\documentclass[a4paper,12pt,titlepageepsf,epsfig,rotating,12pt,a4]{article}
\usepackage{lmodern}
\usepackage[dvipdfmx]{graphicx}
\usepackage{bmpsize}
\usepackage{float}
\usepackage{amsmath}
\usepackage{multicol}
\usepackage{lipsum}
\usepackage[center,small]{caption}
\usepackage{color}
\usepackage{epsf}
\usepackage{epsfig}
\usepackage{url}
\usepackage{subfig}
\usepackage{lineno}

\begin{document}

\vspace*{0.5cm}
\begin{center}
{\large
{\bf
Fingerprint of Tsallis statistics in cosmic ray showers
}}\\
\vspace*{.5cm}
\normalsize {

       {\bf M. Abrah\~ao$^1$,
       W. G. Dantas$^1$, \\
       R. M. de Almeida$^1$,
       D. R. Gratieri$^1$,  and
        T. J. P. Penna$^{2,3}$},

      EEIMVR, Universidade Federal Fluminense, RJ, Brazil$^{1}$\\
      ICEx, Universidade Federal Fluminense, RJ, Brazil$^{2}$\\
      INCT-SC, National Institute of Science and Technology, CNPq, RJ, Brazil$^{3}$\\
      Correspondence should be addressed to R. M. de Almeida: rmenezes@.id.uff.br
}
\end{center}

\vspace*{0.3cm}


\begin{abstract}
We investigate  the impact of the Tsallis non extensive statistics introduced by intrinsic temperature fluctuations in p-Air ultra high energy interactions on observables of cosmic ray showers, such as the slant depth of the maximum $X_{\rm{max}}$ and the muon number on the ground $N_{\mu}$. The results show that these observables are significantly affected by temperature fluctuations and agree qualitatively with the Heitler model predictions.

\end{abstract}

\section{Introduction}

The Pierre Auger Observatory  \cite{Auger1,Auger2} has led to great discoveries in the field of ultra-high energy cosmic rays (UHECRs) such as the confirmation of a
suppression of the cosmic ray flux at energies above $4 \times 10^{19}$  eV, limits on photon 
 and neutrino fluxes  at ultra-high 
energies and a hint of large scale anisotropies  at energies above 8 EeV. Nevertheless many questions related to these particles
are still open.  Particularly interesting is the behavior of the slant depth of the shower maximum with energy.  Understood in terms of the LHC-tuned shower models,  it suggests a gradual shift to a heavier composition, with a large fraction of protons at $10^{18}$ eV, changing to a heavier composition at $10^{19.5}$ eV \cite{Auger2014}. However, we should interpret this result with caution since measurements
of shower properties performed by the Auger Collaboration have revealed inconsistencies between data and present shower models. For instance, the  Pierre Auger Collaboration has reported the first hybrid measurement of the average muon number in inclined air showers at ultra high energies, suggesting a muon deficit in simulations of about $30\%$ to $80_{-20}^{+17} ({\rm sys}) \%$ at $10^{19}$ eV, depending on the hadronic interaction model \cite{Auger-Muon-Deficit}.   Hence  the measured behavior of the slant depth of the shower maximum evolution could be understood as a hint of a  new hadronic interaction physics at energy scales beyond the reach of LHC.

 In this work we will deal with hadronic interactions in a statistical model, as first introduced by Hagedorn \cite{Hagedorn} ideas in the sixties.  Recently a power-law function based on the Tsallis statistics \cite{Tsallis-Paper} has been widely used on fitting the transverse momentum ($p_T$) and pseudo-rapidity ($\eta$) distributions measured in high-energy collisions 
\cite{tsallis-fit-1, tsallis-fit-2, tsallis-fit-3, tsallis-fit-4, tsallis-fit-5} while several studies have been devoted to discuss these 
results in the literature \cite{tsallis-lit-1, tsallis-lit-2, tsallis-lit-3, tsallis-lit-4, tsallis-lit-5, tsallis-lit-6, tsallis-lit-7,
tsallis-lit-8, tsallis-lit-revisao}.   The Tsallis statistics,  extensively used in different branches of science, is often used to describe systems which display properties like memory effects, long range interactions,  intrinsic fluctuations, (multi)fractal phase 
space and so on. It consists in replacing the classical Boltzmann-Gibbs entropy $(S_{BG})$ by the form proposed by Tsallis

\begin{equation}
S_q = \frac{(1-\sum_i p_i^{q})}{q-1} \  \mathrel{\mathop{\Longrightarrow}^{\mathrm{q \rightarrow 1}}}  \ S_{BG} = - \sum_i p_i \ln p_i,
\end{equation}where $p_i$  is the probability of a particle occupy the state $i$ and $q$ is
 the Tsallis index. This definition comprises the Boltzmann-Gibbs entropy as a particular case, where $q=1$. 
On the other hand, a straight consequence from this expression is that the generalized entropy is no longer 
an extensive quantity, once we can verify that

\begin{equation}
S_q(A+B) = S_q(A) + S_q(B) + (1-q) S_q(A)S_q(B),
\end{equation} with the parameter $q$ being a measure of the nonextensivity of the system. As a consequence, we must replace the usual exponential Boltzmann-Gibbs distribution, $\exp(-E/T)$, by the Tsallis power-law distribution

\begin{equation} \label{expq}
f(E)= \frac{(2-q)}{T}\left[ 1 - (1-q) \frac{E}{T} \right]^{\frac{1}{1-q}},
\end{equation} where $E$ is the state energy and $T$ is the temperature of the system.

According to \cite{tsallis-lit-revisao}, the behaviors presented by the transverse momentum and pseudo-rapidity distributions, in high-energies domain,
are best described using a nonexponential distribution, such as the one proposed by Tsallis. In fact, following the ideas discussed
in \cite{tsallis-lit-revisao}, that behavior emerges from fluctuations of the thermal energy within the gas of quarks and gluons before the hadronization process.  Using this approach, we can relate the parameter $q$ with those thermal fluctuations, 

\begin{equation}
q = 1 + \frac{\sigma^{2}_{T}}{\left< T \right>^2}  = 1 +  \frac{\left< \left( \frac{1}{T}\right)^2 \right> - \left<  \left( \frac{1}{T}\right) \right>^2 }{\left<  \left( \frac{1}{T}\right) \right>^2}.
\end{equation}Obviously, when $q=1$ we recover the expected result obtained in the Boltzmann-Gibbs description, where we get an
equilibrium at temperature $T$.

\begin{figure}[!htp!]
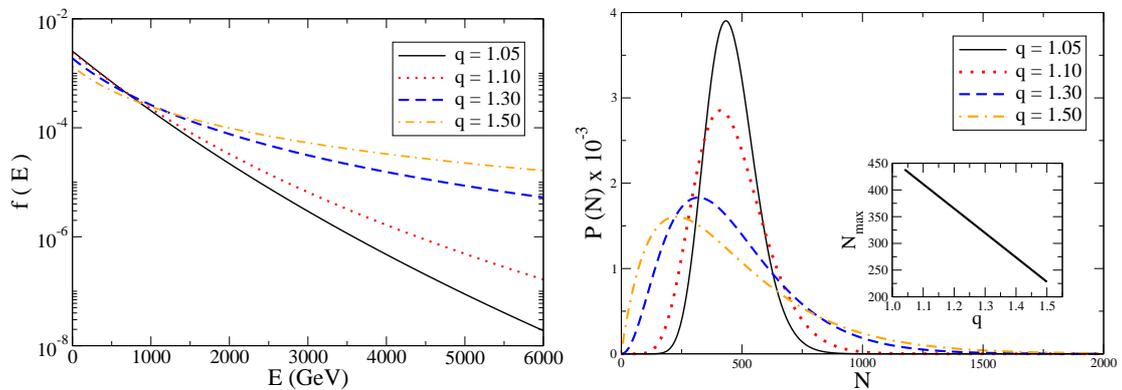

\begin{center}
\includegraphics[totalheight=2.0in, angle=0]{grafico2.eps}
\includegraphics[totalheight=2.0in, angle=0]{histos.eps}
\caption{Tsallis energy distributions (left panel) and the corresponding multiplicity distributions (right panel) as a function of $q$  for a fixed value of temperature $T$.}
\label{histo}
\end{center}
\end{figure}

By assuming such scenario in which the temperature $T$ fluctuates within each collision,   the energy distribution of the particles generated in a single high energy interaction follows a power law Tsallis distribution, given by eq. \ref{expq}. The left panel of Figure \ref{histo} presents the Tsallis energy distribution $f(E)$ with a fixed temperature $T$ for different values of $q$. We can see that higher the $q$ values, the probability for generating particles with larger energy values become greater. As a consequence of the
total energy conservation constraint, $\sum_{i=1}^{i=N} E_i = E_{CM}$, where $E_{CM}$ is the total energy of the interaction in the center of momentum frame, it can be shown that the Tsallis statistics  leads to a Negative Binomial multiplicity distribution given by

\begin{equation} \label{P_N}
P(N) = \frac{(q-1)^N}{N!}  \frac{q-1}{2-q}  \frac{\Gamma(N+1+\frac{2-q}{q-1})}{\Gamma(\frac{2-q}{q-1})}  \left( \frac{E}{T_L}\right)^N \left[   1 - (1-q) \frac{E}{T} \right]^{-N + \frac{1}{1-q}}.
\end{equation} Such distribution has a form shown in the right panel of the Figure \ref{histo}, where it is possible to see how its maximum is affected by $q$, becoming closer to zero as $q$ grows. Besides, the inset plot of this figure shows that the relationship between the value for the maximum of the multiplicity distribution and $q$ is quite linear, at least in that domain of $q$ values.  Therefore, one can see that the introduction of the Tsallis statistics in this context changes the energy, momenta and multiplicity distributions of the particles generated in the hadronic interaction. 

The transverse momentum and pseudo-rapidity distributions resulting from high energy collisions measured by several experiments show a large discrepancy in the values of the parameter $T$, reflecting different physics for the transverse and the longitudinal space. The transverse distributions are thermal-like, presenting a parameter $T_T$ almost independent of the energy while those from the longitudinal space have a temperature  sensitive to the energy of the collision, understood as the mean energy available per produced particle \cite{tsallis-lit-revisao}, $T_L = k E_{CM}/\left< N \right>$, where $\left< N \right>$ and $k$  are, respectivelly,  the mean multiplicity and inelasticity of the interaction. Moreover, since the measured Tsallis index for the longitudinal space $q_L$ is much larger than measured for the transverse space $q_T$ and $T_L \gg T_T$, the resulting $q \sim q_L$.  Also, as verified by simulations, the transverse momentum distribution has a minor contribution on the cosmic ray observables studied in this work.  Therefore, from now on, we will assume a statistical equilibrium for the transverse momentum space and we will refer to the entropic index $q = q_L$ and temperature $T = T_L$. 

The goal of the present paper is to study the impact of the temperature $T$ fluctuations, represented by the parameter $q$, on the shower maximum, $X_{\rm{max}}$, and number of muons on the ground, $N_{\mu}$. The simulations performed in this work are described in section \ref{Simulations}. In section \ref{Results} we present the results of the simulations and discuss them in light of the Heitler Model. Finally, we present the conclusions of this work in section \ref{Conclusions}.

\section{Simulations} \label{Simulations}

For all simulations presented in this work, we have used CORSIKA 7.40 \cite{CORSIKA} with the 
interaction models Sibyll 2.1 \cite{Ahn:2009wx} and GHEISHA 2002d \cite{gheisha}, for high and low 
energy processes, respectively. The muon energy threshold used in the simulations is 0.3 GeV and the array detector position is at 1400 m above sea level, corresponding to the mean altitude of the Pierre Auger Observatory. The air shower simulation chain is as follows: first we simulate the secondaries generated in the collision between a cosmic ray and a nucleus of the upper atmosphere externally by assuming that the hadronization process is described by the Tsallis statistics;  the resulting particle list is 
then inserted back into CORSIKA (using the stacking option and sampling option with thinning = $10^{-6}$) to proceed with usual cascade development through the atmosphere.  Such
a procedure was performed 1000 times for each of several values of $q$ (1.01, 1.025, 1.05, 1.075, 1.10, 1.15, 1.20, 1.25, 1.30, 1.35, 1.40, 1.45 and 1.50)  for showers with zenith angle $\theta = 38^o$ initiated by a  proton of fixed energy $E = 10^{18}$ eV. The reason for limiting the entropic index to $q = 1.5$  in this work is that the mean value of the Tsallis distribution $f(E)$, given by $\left< E \right > = T/(3 - 2q)$, is well defined only for $1 \leq q \leq 1.5$ \cite{tsallis-lit-5}. This model assumes that the Sibyll predictions are valid for lower energies since they are tuned by accelerator data while it fails for higher energies. This added to the parametrizations of the LHC transverse momentum and pseudo-rapidity distributions by the Tsallis statistics justifies its use in this work for energies above  $E \sim 10^{18}$ eV. The point of the air shower first interaction is determined using the $p$-Air cross-section predicted by the Epos 1.99 model \cite{Epos}. The reason for using this value instead of the one predicted by the Sibyll model is that the later presents a large discrepancy in relation to that measured by the Pierre Auger Collaboration \cite{AugerCrossSection}. The mean multiplicity $\left< N \right>$ and inelasticity distribution of the p-Air interaction used in this work were extracted from p-Air interaction simulations using the Sibyll model.  Since the Tsallis distribution is non-extensive, generating particle energies $E_i$ according to this distribution subject to the constraint  $\sum_{i=1}^{i=N} E_i = E_{CM}$ is not a simple task because the  probabilities $f(E_i)$ associated to each particle do not factorize \cite{c_beck_escort}.  Therefore, we perform the simulation process according the following procedure: first, we select the number of particles generated in the p-Air interaction using the $P(N)$ expression given by eq. \ref{P_N} and we assign an energy $E_i$ to each particle $i$ according to the Tsallis distribution.  Then two particles $i$ and $j$  are randomly selected and a random fraction $\Delta E_i$ of the energy $E_i$ is given to particle $j$ in such a way that the new values of energies are $E_{i,new} = E_{i} - \Delta E_{i}$ and $E_{j,new} = E_{j} + \Delta E_{i}$. After that, we compute the deviation of this energy distribution in relation to Tsallis distribution $f(E)$ using the $D^2$ estimator defined by:

\begin{equation}
D^2 = \sum_{i=1}^{
N_{bins}} \left(\frac{dN}{dE} - f(E) \right)^2.
\end{equation} If the new $D^2$ value is smaller than the previous one, we accept the changes  in energy of the particles $i$ and $j$, otherwise we cancel them. We keep repeating this procedure for another pair of particles. The whole process continues until the $D^2$ value is stabilized.  Generally, it takes $10^4$ iterations to reach such stabilization. To be conservative, the simulations presented in this work were performed using $10^5$ iterations for each p-Air interaction. Since we verified through simulations that $X_{\rm{max}}$  and $N_{\mu}$ are not sensitive to changes in $q_T$, all simulations corresponding to the transverse space presented in this work were evaluated assuming a statistical equilibrium, i.e, the Hagedorn \cite{Hagedorn} transverse momentum distribution:

\begin{equation}
\frac{d N}{dp_T} \simeq c p_T \exp \left( - \frac{p_T}{T_T} \right),
\end{equation} with $\left< T_T \right>= 133$ MeV.The type of particles are randomly generated according to Sibyll predictions and once we have generated the particles masses $m$, the longitudinal momentum is obtained as $p_L = \sqrt{E^2 - m^2 - p^2_T}$.  These kinematic variables along with the species of particle complete all the information we need to reintroduce in CORSIKA to proceed with shower propagation through the atmosphere.

\section{Results and discussion} \label{Results}

In the following we describe the impact of $T_L$ fluctuations, represented by the parameter $q = q_L$, on the shower maximum, $X_{\rm{max}}$, and number of muons on the ground. We will discuss them in terms of the predictions of the Heitler model \cite{Heitler, Matthews} as well as of the results achieved in reference \cite{Ulrich_Engel_Unger}. Although extremely simple, the predictions of the Heitler model are remarkable. It assumes that the shower maximum is reached when the energies of particles become smaller than a critical energy, in which energy loss processes dominate the production of new particles in the case of electromagnetic component, or the charged pion interaction length becomes larger than the decay length of pions in muons, in the case of the hadronic one. As a consequence, it predicts an increase of the $\left<X_{\rm{max}}\right>$ for smaller mean multiplicities, since larger multiplicities correspond to lower energy per particle.  Besides, reference \cite{Ulrich_Engel_Unger} describes a detailed investigation of the impact of the multiplicity, hadronic particle production cross section, elasticity and pion charge-ratio on air shower observables with most of the predictions qualitatively understood within the simple Heitler model and its extension to hadronic component. 
 
\begin{figure*}
\center{
\includegraphics[totalheight=2.2in, angle=0]{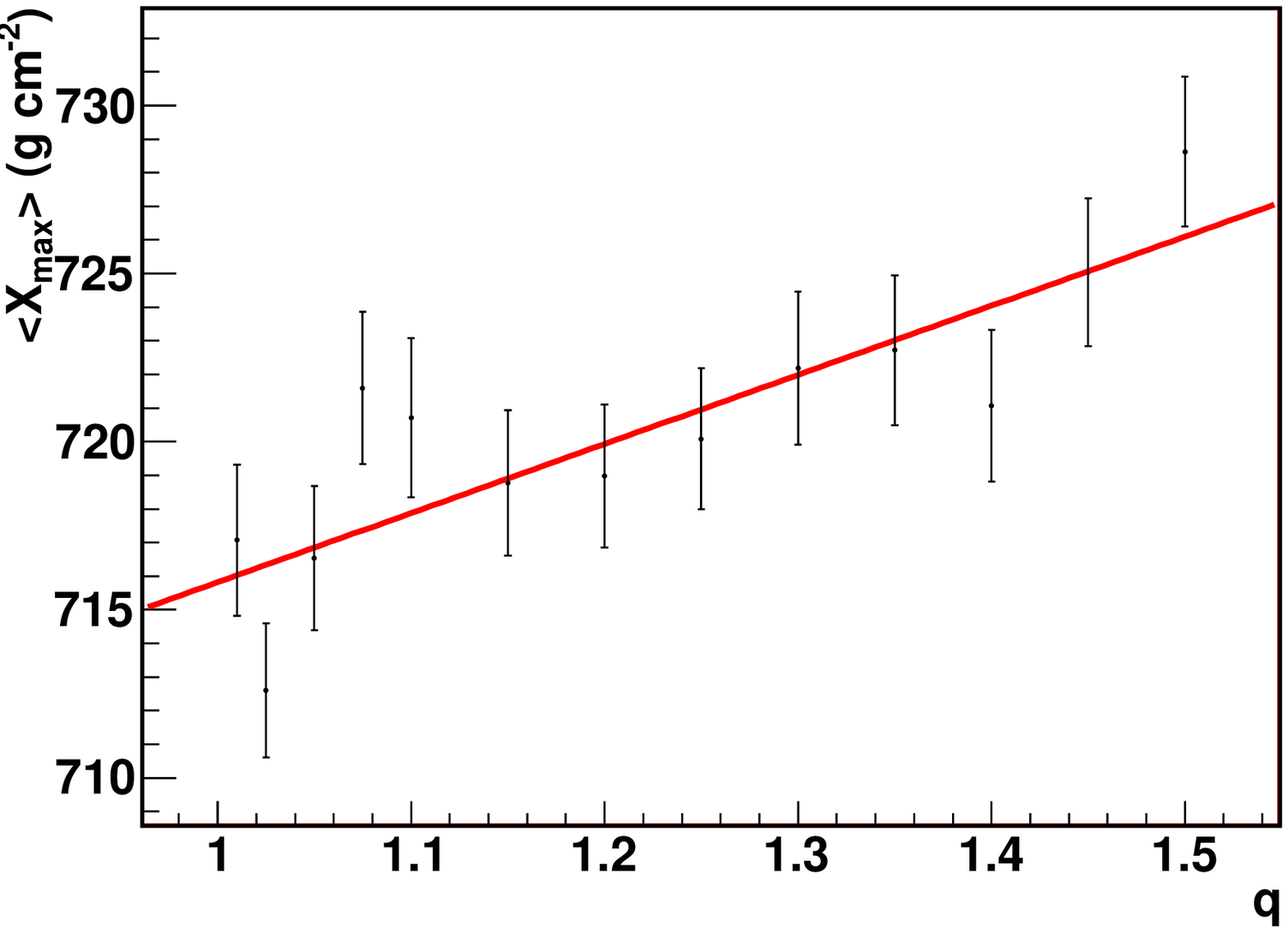}
\includegraphics[totalheight=2.2in, angle=0]{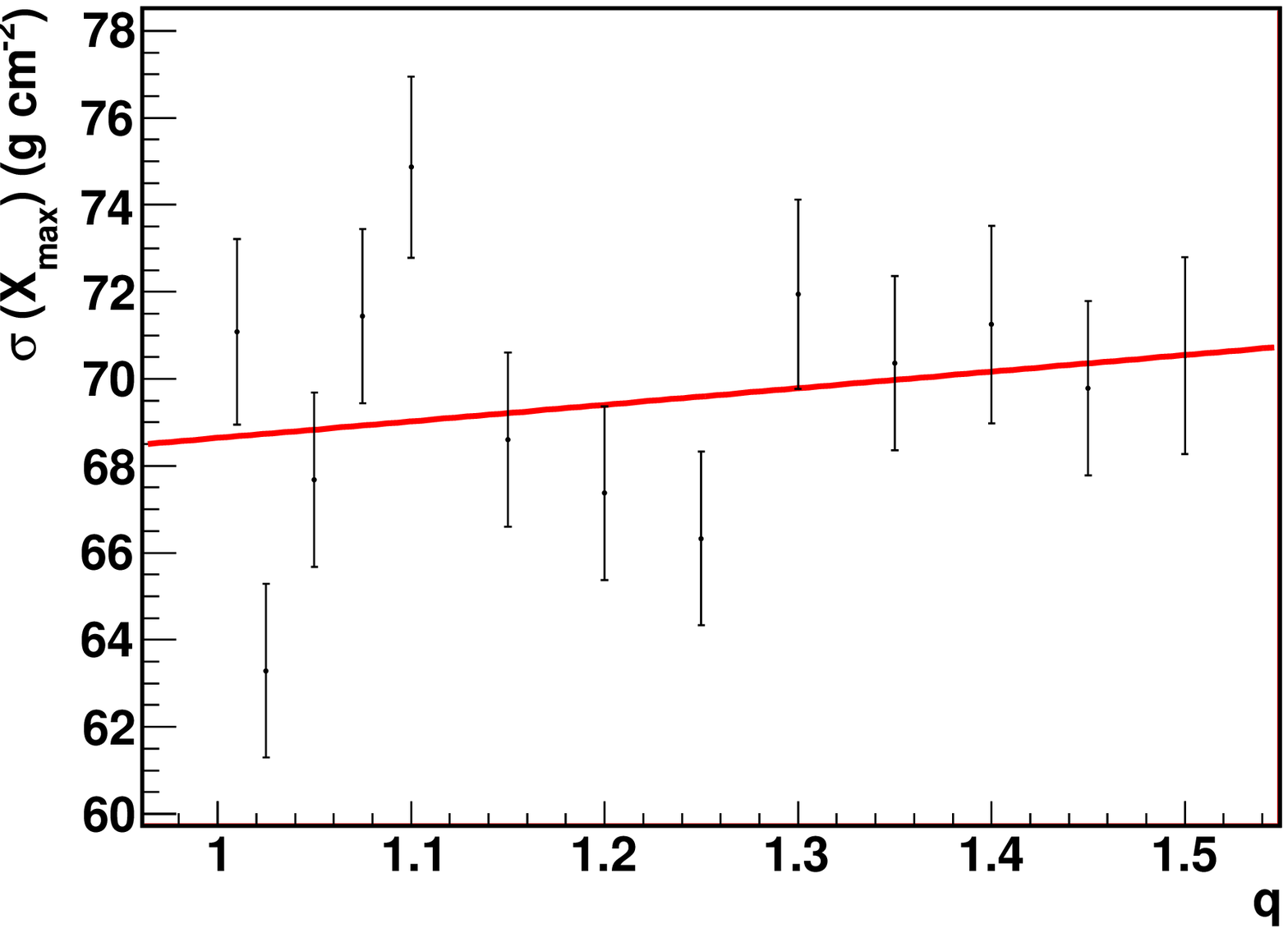}}
\caption{$X_{\rm{max}}$ (left) and $\sigma ({X_{\rm{max}}})$ (right) distributions obtained from air shower simulations initiated by the interaction between a proton of $E = 10^{18}$ eV and a nucleus of the upper atmosphere in which the hadronization process is described by the Tsallis statistical.}
\label{dist_xmax}
\end{figure*}

The Pearson coefficient $\rho$ was used to assess the degree of correlation between air showers observables and $q$. It is defined by

\begin{equation}
\rho = \frac{{\rm cov}(X,Y)}{\sigma(X) \cdot \sigma(Y)}
\end{equation} and measures the linear correlation between two variables $X$ and $Y$, yielding  a value in the interval $\left[ -1, +1 \right]$, with $1$ meaning total positive correlation, $0$  no correlation, and $-1$  total negative correlation. cov(X) is the covariance between $X$ and $Y$ and $\sigma(X)$ and  $\sigma(Y)$ are the standard deviations of variables $X$ and $Y$. The results for the mean depth of shower maximum, $\left<X_{\rm{max}}\right>$, and the fluctuations of $X_{\rm{max}}$ are summarized in Figure \ref{dist_xmax}.  

The left panel shows the $\left<X_{\rm{max}}\right>$ as a function of $q$. A strong correlation is observed yielding  a Pearson correlation coefficient  $\rho = 0.84$. The red line presents the best linear fit corresponding to a $\chi^2_{\nu} = 1.09$.
The comparison between the predictions from Heitler model and reference \cite{Ulrich_Engel_Unger} with our results requires caution since we did not change the mean multiplicity in our simulations of the first interaction. However, the changes in distributions of energy and momenta of the particles generated in the first interaction results in a spread of the multiplicity distribution and a shift of its peak to lower values as it is shown in Figure \ref{histo}. The reason for the strong correlation is that most of the showers generated with larger $q$ values are initiated with smaller multiplicities, or equivalently, with  higher energy per particle. 

Besides, the right panel of Figure \ref{dist_xmax} presents the corresponding plot for the $\sigma (X_{\rm{max}})$ as a function of $q$.  In this case, the observed correlation is weak, with $\rho = 0.21$ and  $\chi^2_{\nu} = 1.84$ corresponding to the best linear fit, shown by the red line. According to the Heitler model, the variance of the ${X_{\rm{max}}}$ distribution depends on the hadronic cross-section interaction, $\sigma_I$, and multiplicity, $N$, via ${\rm V} ({X_{\rm{max}}}) \propto 1/\sigma_{I}^{2} + \ln (2 X_0)^2 {\rm V} (\ln N)$, where $X_0 \sim 37$ g/cm$^2$ is the electromagnetic radiation lenght. In this work, the hadronic interaction is given by the one corresponding to the interaction between a proton and a nucleus of the atmosphere, $\sigma_I = \sigma_{p-Air}$. Although the ${\rm V} (\ln N)$ increases with $q$, the observed correlation between ${\rm V} ({X_{\rm{max}}})$ and $q$ is weak since the increase of spread of the multiplicity distribution for larger $q$ values is dominated by the first term contribution in ${\rm V} ({X_{\rm{max}}})$, as a consequence of the relatively small value of the p-Air cross-section. 

\begin{figure*}
\center{
\includegraphics[totalheight=2.22in, angle=0]{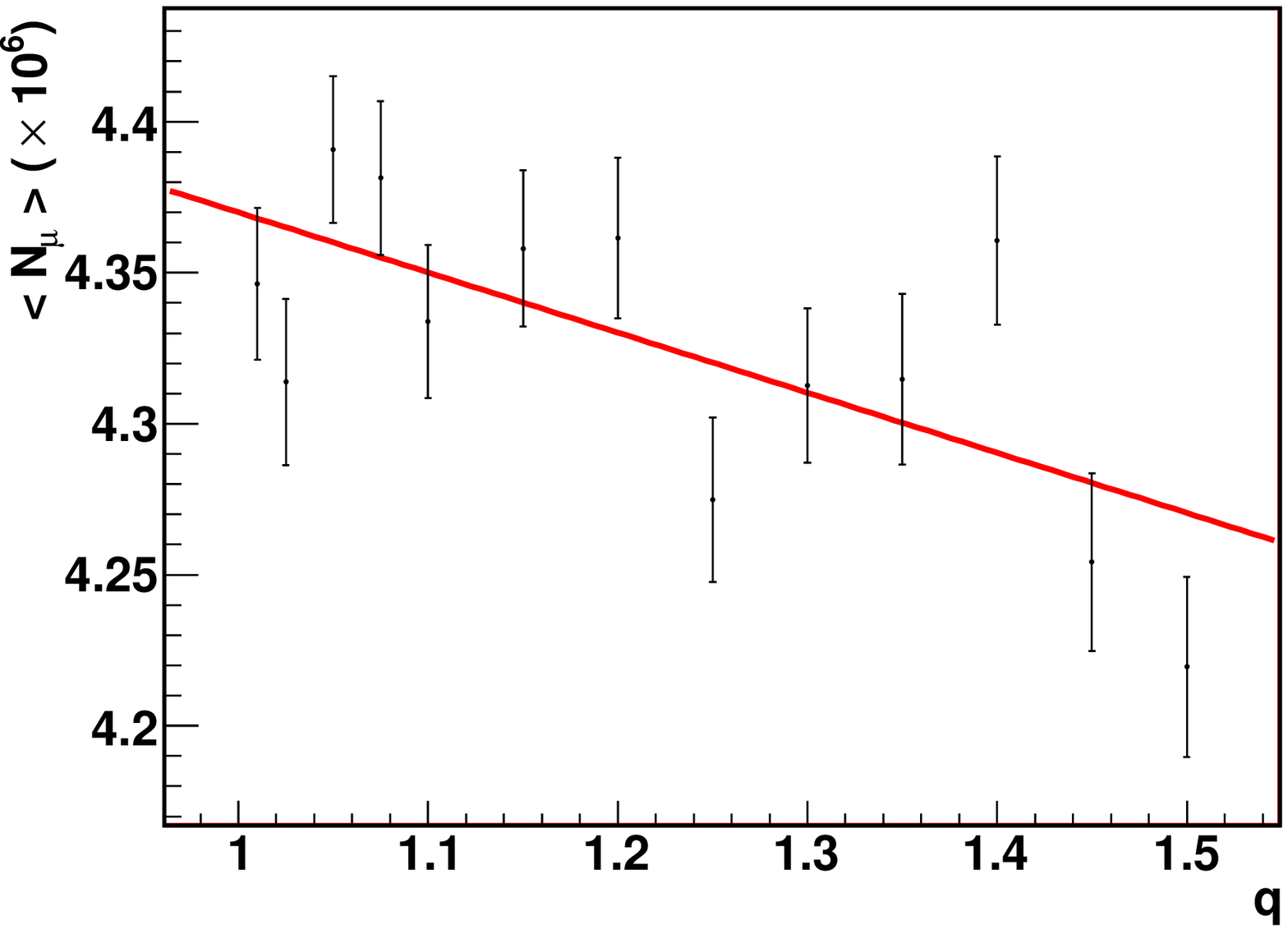}
\includegraphics[totalheight=2.22in, angle=0]{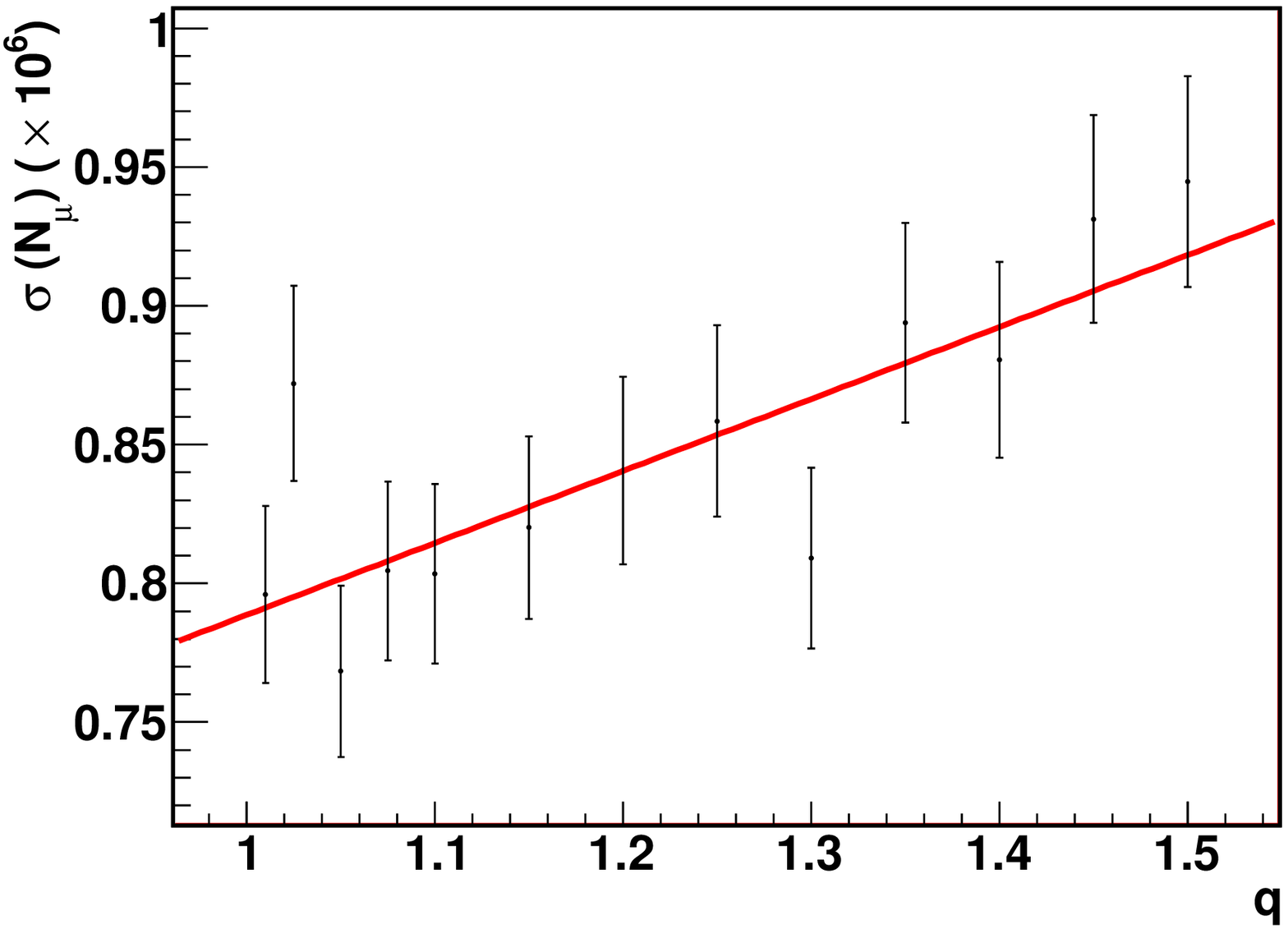}}
\caption{$<N_{\mu}>$ (left) and $\sigma_{N_{\mu}}$ (right) distributions obtained from air shower simulations initiated by the interaction between a proton of $E = 10^{18}$ eV and a nucleus of the upper atmosphere in which the hadronization process is described by the Tsallis statistical.}
\label{dist_mu}
\end{figure*}

The impact of $q$ on the mean number of muons in the ground, $\left<N_{\mu}\right>$, and the fluctuations of $N_{\mu}$ are summarized in Figure \ref{dist_mu}. The  $\left<N_{\mu}\right>$ as a function of $q$, shown in the left panel,  presents a strong anti-correlation with $q$, with $\rho = -0.91$ and $\chi^2_{\nu} = 0.72$ corresponding to the best linear fit, marked in red.  A superficial analysis of the left panels of Figures \ref{dist_xmax} and \ref{dist_mu} could indicate wrongly that $\left<N_{\mu}\right>$ and  $\left<X_{\rm{max}}\right>$ are anti-correlated.  The positive correlation between $\left<N_{\mu}\right>$ and  $\left<X_{\rm{max}}\right>$ position exists but it is weak since muons are hardly attenuated in the atmosphere. Therefore it is not the most important factor for the $\left<N_{\mu}\right>$ behavior as a function of $q$. Indeed muons are mainly produced as a result of pion decay and their abundance in the ground, especially considering the most energetic particles, is strongly correlated with the number of pions in the shower. 
As a consequence of the reduction of the peak of the multiplicity distribution for larger $q$ values, most showers presents lower production of pions in the first interaction, constituting the main reason for the observed anti-correlation between 
$\left<N_{\mu}\right>$ and $q$.  

On the other hand, the right panel of Figure \ref{dist_mu} shows a strong correlation of $\sigma (N_{\mu})$ and $q$, with $\rho = 0.88$, as a natural consequence of the spread of the multiplicity distribution. The red line shows the best linear fit with corresponding $\chi^2_{\nu} = 0.52$.

Finally, Table \ref{summary_results} summarizes the correlation coefficients $\rho$ between air shower observables and $q$ obtained in this work. 

\begin{table}[!htp!]
\begin{center}
\begin{tabular}{|c|c|c|c|c|}\hline
   & $\left< X_{\rm{max}} \right>$  & $\sigma (X_{\rm{max}})$  &  $\left< N_{\mu} \right>$ &  $\sigma (N_{\mu})$ \\ \hline
  $q$ & $\rho = 0.84$  & $\rho = 0.21$  &  $\rho = -0.91$ &  $\rho = 0.88$ \\ \hline
\end{tabular}
\caption{Summary of the correlation coefficients between air shower observables and $q$ obtained in this work.}
\label{summary_results}
\end{center}
\end{table}

\section{Conclusions} \label{Conclusions}

Although the simulations presented in this work are a very simple description of ultra high energy interactions, the results presented here show that intrinsic fluctuations of the system with respect to $T_L$,  given by the parameter $q = q_L$,  change the energy, momenta and multiplicity distributions of the particles generated in the interaction between a cosmic ray and a nucleus of the atmosphere, impacting air shower observables such as the slant depth of the maximum $X_{\rm{max}}$ and the muon number on the ground $N_{\mu}$.  The results show that the higher the temperature fluctuations, the greater the values of the mean slant depth of maximum $\left< X_{\rm{max}}\right>$ and  variance of the number of muons on  the ground $\sigma(N_{\mu})$, with Pearson correlation coefficients of $\rho = 0.84$ and $\rho = 0.88$, respectively. This results from  the spread and shift of the maximum of the multiplicity distribution to lower values for larger temperature fluctuations. Besides, as muons are mainly produced by the decay of charged pions and the shift in the peak of multiplicity distribution reduces the number of such particles generated in the first interaction, the  mean number of muons on the ground $\left< N_{\mu}\right>$ presents a strong negative correlation with $q$, producing a $\rho = -0.91$. On the other hand, the variance of the slant depth distribution $\sigma(X_{\rm{max}})$ presents a very weak correlation with the temperature fluctuations, with $\rho = 0.21$,  because the contribution of the hadronic p-Air interaction fluctuations dominates the one originated from the multiplicity distribution. These results agree qualitatively with the Heitler model and reference \cite{Ulrich_Engel_Unger} predictions. Although these results have been obtained for a specific non-extensive hadronic interaction model, we believe that it  captures the essential features related to the presence of Tsallis statistics in UHECR showers and can shed light to the understanding of their properties as well as of  particle interactions at these energies. Studies regarding different non-extensive particle interaction models are out of the scope of this work and will be addressed in a future work. 

\newpage

\section{Acknowledgments}

We thank Professor J. C.  Anjos for useful discussions in the beginning of this work. This work was partially supported by the ``Conselho Nacional de Desenvolvimento Cient\'\i fico e Tecnol\'ogico" (CNPq) and by the ``Funda\c c\~ao Carlos Chagas Filho de Amparo \`a Pesquisa do Estado do Rio de Janeiro" (FAPERJ).

\end{document}